\newcommand{\lam}{\lambda}

\documentclass[12pt]{article}

\usepackage{latexsym}

\begin{document}

\title{On the scalar sector of the covariant graviton
two-point function in de\ Sitter spacetime}

\author{Atsushi Higuchi$^1$ and Spyros S.\ Kouris$^2$\\ 
{\normalsize Department of Mathematics, University of York}\\ 
{\normalsize Heslington, York, YO10 5DD, United Kingdom}\\
{\normalsize $^1$Email: ah28@york.ac.uk}\\
{\normalsize $^2$Email: ssk101@york.ac.uk}}

\date{22 May, 200}
\maketitle

\begin{abstract}
We examine the scalar sector 
of the covariant graviton two-point function in de\ Sitter spacetime.
This sector consists of the pure-trace part and another part described by
a scalar field.  We show that it does not
contribute to two-point functions of gauge-invariant quantities.  We
also demonstrate
that the long-distance growth present in some gauges
is absent in this sector for a wide range of gauge parameters.
\end{abstract}

\section{Introduction}

The covariant graviton two-point function in de\ Sitter spacetime (CGTF)
has been 
studied by several authors~\cite{Allen86:4}--\cite{Takook}.  (See, e.g.,
\cite{HawkEllis} for a description of de\ Sitter spacetime.)
It is known that the CGTF grows with the distance
between the two points in some gauges~\cite{AllenTuryn,AM}, 
although infrared
divergences have been shown to be absent~\cite{Allen86:4}.
In particular, it is known that the pure-trace part of the CGTF grows with
the distance in a gauge where
the traceless part is 
divergence-free~\cite{AM}.  
(We call this gauge the Landau gauge in this paper.) 

Recently, a non-covariant physical
graviton two-point function, which contains only the two physical 
polarizations, was computed
in open de\ Sitter spacetime~\cite{HHT}.  
It was found that this two-point function does not grow as the
distance between the two points increases.  Subsequently, the present authors
showed~\cite{HigKou} 
that the logarithmic growth of the non-covariant physical 
two-point function in spatially-flat de\ Sitter spacetime~\cite{Allen87:1}
is a gauge artefact.
These results 
suggest that the seemingly problematic
long-distance behaviour in the pure-trace part of the CGTF 
may also be a gauge artefact.  Interestingly, the one-loop effect
in pure-trace external gravitational fields has been shown to 
vanish~\cite{AIT}.
In fact one of the present authors (AH) showed
under some assumptions~\cite{Higuchi87:1}
that the contribution from the pure-trace part and 
that from another scalar part of
the CGTF together take 
a pure-gauge form for a one-parameter family of gauges
which includes the 
Landau gauge as 
a limit.\footnote{Allen and Turyn~\cite{AllenTuryn} have also pointed 
out that the terms
which grow at large distances in these two parts of the CGTF
are in a pure-gauge form
in a particular gauge in the Euclidean approach.} (We will call the sector
consisting of these two
parts the scalar sector.)  
In this paper we prove this
fact without making any assumptions, thus confirming that the scalar sector
of the 
CGTF does not contribute to two-point functions of physical quantities
at the tree level. 
We also
emphasize that the mass of the scalar sector is gauge dependent and can
be chosen to be a value for which there is no long-distance growth. 

The rest of the paper is organized as follows.
In section 2 we recall some essential facts about
the scalar field theory in de\ Sitter spacetime
as a preliminary.
In section 3 we examine the field equations
of linearized gravity in a one-parameter family of covariant gauges
and identify the scalar sector, which will be the subject 
of this paper.
In section 4 we find some commutator functions, which will be
used to find the scalar-sector two-point function.
In section 5 we present our main result, i.e.
the scalar sector of the CGTF, and make some remarks
about the full CGTF.
In Appendix A an explicit form of the scalar sector
two-point function is presented for a particular value of the gauge parameter,
where some simplification occurs.

The metric signature is $(-+++)$ and we set $c = \hbar = 1$ throughout
this paper. 

\section{Scalar field theory}

Most of the discussions in this paper are independent of the explicit metric,
but for definiteness we work mainly 
with the metric 
\begin{equation}
ds^2= -dt^2 + H^{-2}\cosh^2 Ht\,dS_3^2\,,
\label{metric2}
\end{equation}
where $dS_3^2$ is the line element of the unit 3-sphere.
The Hubble constant is $H$ and the cosmological constant is $3H^2$ here.

Let us consider the scalar field theory with the Lagrangian density
\begin{equation}
{\cal L}_S = -\frac{\sqrt{-g}}{2}\left[
\nabla_a \phi\nabla^a \phi + \frac{12H^2}{\alpha-3}\phi^2\right]\,,
\label{scalar}
\end{equation}
where $\nabla_a$ is the covariant derivative in
the background de\ Sitter spacetime. Indices are raised and lowered
by the de\ Sitter metric $g_{ab}$ given by (\ref{metric2}).
The mass parameter, $12H^2/(\alpha-3)$,
is given in a form which will be useful later.\footnote{The theory
may not be well defined for some values
of $\alpha$.  We avoid these values.}  The corresponding Euler-Lagrange
equation is 
\begin{equation}
L^{(S)}\phi \equiv \left( \Box - \frac{12H^2}{\alpha-3}\right)\phi = 0\,,
\label{scELeq}
\end{equation}
where $\Box \equiv \nabla_a\nabla^a$.
We expand the field $\phi$ as
\begin{equation}
\phi = \sum_{l\sigma}\left( a_{l\sigma}\phi^{(l\sigma)} 
+ a_{l\sigma}^\dagger \overline{\phi^{(l\sigma)}} \right)\,,
\end{equation}
where the mode functions $\phi^{(l\sigma)}$ are given by
\begin{equation}
\phi^{(l\sigma)} = f_l(t)Y_{l\sigma}  \label{scmode}
\end{equation}
with the $Y_{l\sigma}$ being  spherical harmonics on the unit 3-sphere
with angular momentum $l$.  They satisfy
$\Delta Y_{l\sigma} = -l(l+2)Y_{l\sigma}$, where $\Delta$ is
the Laplace operator on the unit 3-sphere. (See, e.g., \cite{ChodosM} for a
concise description of spherical harmonics in higher dimensions.)  
The label $\sigma$ distinguishes
the spherical harmonics with the same value of $l$.  We require
\begin{equation}
\int_{S^3} d\Omega\, \overline{Y_{l\sigma}}Y_{l'\sigma'}
= \delta_{ll'}\delta_{\sigma\sigma'}\,. \label{SPNM}
\end{equation}
The functions $f_l(t)$ satisfy
\begin{equation}
\left[ \frac{d^2\ }{d t^2}+
 3H\tanh Ht\frac{d\ }{dt} + \frac{l(l+2)H^2}{\cosh^2 Ht}
+ \frac{12H^2}{\alpha-3}\right]f_{l}(t) = 0\,.
\end{equation} 
We normalize the functions $f_l(t)$ by requiring the following Wronskian: 
\begin{equation}
W(\overline{f_l}, f_l) \equiv
\overline{f_l} \frac{df_l}{dt} - f_l \frac{d\overline{f_l}}{dt}
=-\frac{iH^3}{\cosh^3 Ht}\,. \label{Wronskian}
\end{equation}

Given any two solutions $\phi^{(1)}$ and $\phi^{(2)}$,
the following current is conserved:
\begin{equation}
{\cal J}^{c}(\phi^{(1)},\phi^{(2)}) =
\phi^{(2)}\nabla^c \phi^{(1)}-\phi^{(1)}\nabla^c\phi^{(2)}\,.
\end{equation}
We define
the Klein-Gordon inner product of 
$\phi^{(1)}$ and $\phi^{(2)}$ as
\begin{equation}
(\phi^{(1)}|\phi^{(2)})
\equiv i\int_{\Sigma} d\Sigma_c {\cal J}^{c}(\overline{\phi^{(1)}},
\phi^{(2)})\,,
\end{equation}
where $\Sigma$ is any Cauchy surface and where
$d\Sigma_c = d\Sigma\,n_c$ with $n_c$ being the future-pointing unit normal
to $\Sigma$. (This can be generalized to fields of higher spin. See, e.g.
\cite{Friedman}.)
Here, $d\Sigma = d\theta_1 d\theta_2 d\theta_3\sqrt{\eta}$, where 
the $\theta_i$ are the coordinates on $\Sigma$ and $\eta$ is the determinant
of the metric on $\Sigma$.
The product $(\phi^{(1)}|\phi^{(2)})$ is independent
of the Cauchy surface $\Sigma$
because the current ${\cal J}^{c}$ is conserved.
We find using (\ref{SPNM}) and (\ref{Wronskian})
\begin{eqnarray}
(\phi^{(l\sigma)}|\phi^{(l'\sigma')}) 
& = & \delta_{ll'}\delta_{\sigma\sigma'}\,,
  \\
(\overline{\phi^{(l\sigma)}}|\phi^{(l'\sigma')}) & = & 0\,.
\end{eqnarray}
The usual canonical quantization leads to the commutation relations
\begin{equation}
\left[ a_{l\sigma}, a_{l'\sigma'}^\dagger\right]
= \delta_{ll'}\delta_{\sigma\sigma'}  \label{sccommut}
\end{equation}
with all other commutators vanishing. 
We define the vacuum state $|0\rangle$
by requiring $a_{l\sigma}|0\rangle = 0$ for all $l$ and $\sigma$.
There is some freedom in choosing $f_l(t)$ and, consequently, in choosing
the vacuum state.  The standard choice leads 
to the so-called Euclidean~\cite{GibHawk} (or Bunch-Davies~\cite{BD}) vacuum, 
which is invariant under
the de\ Sitter group.  We choose this
vacuum, but the explicit form 
of $f_l(t)$ is not necessary here
(see, e.g., \cite{Tagirov}).

The two-point function of the scalar field $\phi$ is expressed as
\begin{eqnarray}
\Delta_{+}(x,x') & \equiv & \langle 0|\phi(x)\phi(x')|0\rangle  \nonumber \\
 & = & \sum_{l\sigma}\phi^{(l\sigma)}(x)\overline{\phi^{(l\sigma)}(x')}\,,
\label{sctwo-pt}
\end{eqnarray}
where $x$ and $x'$ are spacetime points. 
The function $\Delta_{+}(x,x')$ has been calculated by several 
authors~\cite{BD,Spindel,AllenJacobson}.
For example, Allen and Jacobson~\cite{AllenJacobson}
give it as
\begin{equation}
\Delta_{+}(x,x') =\frac{\Gamma(a_+)\Gamma(a_-)H^2}{16\pi^2}F(a_+,a_-;2;z)\,,
\label{Bruce}
\end{equation}
where 
\begin{eqnarray}
a_{\pm}& = & \frac{3}{2}\pm \left(\frac{9}{4}-\frac{M^2}{H^2}\right)^{1/2}
 \ \ 
{\rm with}\ \ 
M^2  = \frac{12H^2}{\alpha-3}\,, \\
z & = & \cos^2\left(\frac{\mu(x,x')H}{2}\right)\,. \label{defcos}
\end{eqnarray}
Here, the function $\mu(x,x')$ is the spacelike geodesic distance between
points 
$x$ and $x'$, and $F(a,b;c;z)$ is the hypergeometric function.  
The variable $z$ can be extended to the case where there is
no spacelike geodesic between $x$ and $x'$.  If we write the de\ Sitter
metric as
\begin{equation}
ds^2 = \frac{1}{H^2\lambda^2}(-d\lambda^2 + d{\bf x}^2)\,,
\label{metric0}
\end{equation}
with ${\bf x} = (x_1, x_2, x_3)$, 
then for $x = (\lambda, {\bf x})$ and 
$x'=(\lambda',{\bf x}')$ we have
\begin{equation}
z = \frac{(\lambda+\lambda')^2 - \|{\bf x}-{\bf x}'\|^2}{4\lambda\lambda'} \,.
\label{alternative}
\end{equation}
(This can readily be inferred by comparing the two-point functions in
\cite{BD} and \cite{AllenJacobson}.)
Hence, the large-distance limit corresponds to the limit $z\to -\infty$.  The 
large-distance limit of (\ref{Bruce}), which will be useful later, can be found
as $|\Delta_+(x,x')| \sim (-z)^{-a_{-}}$ if $0 < M^2 \leq 9/4$ and
$|\Delta_+(x,x')| \sim (-z)^{-3/2}$ if $9/4 \leq M^2$, up to a constant factor.
Thus, if $M^2 > 0$, i.e. if $\alpha > 3$, the scalar two-point function
$\Delta_{+}(x,x')$ tends to zero as $z \to -\infty$.

The commutator function, which is often called the Schwinger function, is
\begin{equation}
\left[ \phi(x),\phi(x')\right]
= \Delta_+(x,x') - \Delta_+(x',x)\,.
\end{equation}
Now, define the advanced and retarded Green functions, $G^{(S,A)}(x,x')$
and $G^{(S,R)}(x,x')$, by requiring that
\begin{equation}
L^{(S)}_xG^{(S,A/R)}(x,x') = \delta^{4}(x,x')\,,
\end{equation}
and that $G^{(S,A)}(x,x')$ ($G^{(S,R)}(x,x')$) 
vanish if $x$ is in the future (past)
of $x'$. (The subscript $x$ in $L^{(S)}_x$ indicates that the differential
operator $L^{(S)}$ acts at $x$ rather than at $x'$.) 
We have defined $\delta^4(x,y)$ by
\begin{equation}
\int dV_y\, \delta^4(x,y)f(y) = f(x)
\end{equation}
for any smooth and compactly-supported function $f(x)$, where
\begin{equation}
dV_y \equiv d^4 y \,\sqrt{-g(y)}\,.
\end{equation}
Then, as pointed out by Peierls \cite{Peierls},
the commutator function is given by the advanced-minus-retarded
Green function 
\begin{equation}
E^{(S)}(x,x') \equiv G^{(S,A)}(x,x') - G^{(S,R)}(x,x')
\end{equation}
as 
\begin{equation}
\left[ \phi(x), \phi(x')\right] = i E^{(S)}(x,x')\,.
\end{equation}

\section{The scalar-sector mode functions}

The Lagrangian density for linearized gravity can be chosen as 
\begin {eqnarray}
\mathcal{L}_{\rm inv} = \sqrt{-g}
\left[ \frac{1}{2}\nabla_{a}h^{ac}\nabla^{b}h_{bc}
-\frac{1}{4}\nabla_{a}h_{bc}\nabla^{a}h^{bc}
+\frac{1}{4}(\nabla^{a}h-2\nabla^{b} h^{a}_{\ b})\nabla_{a}h \right.
 \nonumber \\
\left. -\frac{1}{2}H^2\left(
h_{ab}h^{ab}+\frac{1}{2}h^2\right)\right] \label{Lagden}
\end{eqnarray}
with $h = h^{a}_{\ a}$. 
This Lagrangian density is invariant
under the gauge transformation
$$
h_{ab}\to h_{ab}+\nabla_{a}\Lambda_{b}+\nabla_{b}\Lambda_{a}
$$
up to a total divergence.
Therefore, one 
needs to fix the gauge for the canonical quantization of $h_{ab}$.
For this purpose we add the following
gauge-fixing term in the Lagrangian density:
\begin{eqnarray}
\mathcal{L}_{\rm gf}=-\frac{\sqrt{-g}}{2\alpha}
\left( \nabla_a h^{ab}-\frac{1+\beta}{\beta}\nabla^{b}h\right)
\left( \nabla^c h_{cb}-\frac{1+\beta}{\beta}\nabla_{b}h\right)\,.
\end{eqnarray}
Then the Euler-Lagrange field equations derived from  
${\cal L}_{\rm inv}+{\cal L}_{\rm gf}$ are 
\begin{eqnarray}
{L^{(T)}_{ab}}^{cd}h_{cd}
\equiv \frac{1}{2}\Box h_{ab}-\left(\frac{1}{2}-\frac{1}{2\alpha}\right)
\left(\nabla_{a}\nabla_{c} h^{c}_{\ b}
+ \nabla_{b}\nabla_{c} h^{c}_{\ a}\right) \nonumber \\
\ \ \  +\left[\frac{1}{2}
-\frac{\beta+1}{\alpha\beta}\right]
 \nabla_a\nabla_b h+\left[\frac{(\beta+1)^2}{\alpha\beta^2}
-\frac{1}{2}\right]g_{ab}\Box h
  \nonumber \\
\ \ \  + \frac{1}{2}g_{ab}\left(1-\frac{2(1+\beta)}{\alpha\beta}\right)
\nabla_{c}\nabla_{d} h^{cd}
-H^2\left(h_{ab}+\frac{1}{2}g_{ab}h\right) = 0\,.
\label{Eulag1}
\end{eqnarray} 

The scalar sector of the field $h_{ab}$ satisfying this equation
can be extracted as follows~\cite{Higuchi87:1}.
First, by taking the trace of equation (\ref{Eulag1}) we find
\begin{equation}
\left[\frac{4(1+\beta)^2}{\alpha\beta^2}-\frac{1+\beta}{\alpha\beta}
-1\right]\Box h -3H^2h
+\left( 1-\frac{3}{\alpha}-\frac{4}{\alpha\beta}\right)
\nabla^a\nabla^b h_{ab}=0\,.  \label{TraceEq}
\end{equation}
There is mixing between the trace $h$ and the traceless part of 
$h_{ab}$ in general as can be seen from this equation.  
This mixing can be avoided by choosing the parameter
$\beta$ as 
\begin{equation}
\beta=\frac{4}{\alpha-3}\,. \label{beta}
\end{equation}
We make this choice in the rest of this paper.  We also assume that
$\alpha\neq 0,\,3$ though we will consider the limit $\alpha\to 0$
in the concluding section.
Equation (\ref{TraceEq}) now reads
\begin{equation}
\left( \Box-\frac{12H^2}{\alpha-3}\right) h=0\,.
\label{eqnh}
\end{equation}
We define the traceless part $h^{(l)}_{ab}$ of $h_{ab}$ by
\begin{equation}
h^{(l)}_{ab} \equiv h_{ab} - h^{(t)}_{ab}\,, \label{trls}
\end{equation}
where
\begin{equation}
h^{(t)}_{ab} \equiv \frac{1}{4}\,g_{ab}h(x)\,.  \label{trdef}
\end{equation}
By substituting (\ref{trls})
in (\ref{Eulag1}) with the choice (\ref{beta}) we find
\begin{eqnarray}
&&\frac{1}{2}\Box h^{(l)}_{ab}
-\left(\frac{1}{2}-\frac{1}{2\alpha}\right)
\left(\nabla_{a}\nabla_{c} {h^{(l)c}}_{b}
+ \nabla_{b}\nabla_{c} {h^{(l)c}}_{a}\right)
  \nonumber \\
&& + \left( \frac{1}{4}- \frac{1}{4\alpha}\right)
g_{ab}\nabla_{c}\nabla_{d} h^{(l)cd}
-H^2 h^{(l)}_{ab} = 0\, . \label{Eulag2}
\end{eqnarray} 
Next, define a scalar field $B$ by
\begin{equation}
B \equiv 
\frac{(\alpha-3)^2}{36\alpha H^4}\nabla^a \nabla^b h_{ab}^{(l)}\,.
\label{DefB}
\end{equation}
By using the fact that the Riemann tensor takes the form
$R_{abcd} = H^2(g_{ac}g_{bd} - g_{ad}g_{bc})$,
we find
\begin{equation}
\left( \Box-\frac{12H^2}{\alpha-3}\right) B =0\,.
\label{eqnB}
\end{equation}
Note that the two fields $h$ and $B$ satisfy the same equation as the
scalar field $\phi$ discussed in the previous section.
Hence, we can expand these fields as
\begin{eqnarray}
B(x) & = & \sum_{l\sigma}\left[ b_{l\sigma}\phi^{(l\sigma)}(x) + 
b_{l\sigma}^\dagger \overline{\phi^{(l\sigma)}(x)}\right]\,, \label{Bexp} \\
h(x) & = & \sum_{l\sigma}\left[ c_{l\sigma}\phi^{(l\sigma)}(x) + 
c_{l\sigma}^\dagger \overline{\phi^{(l\sigma)}(x)}\right]\,. \label{trexp} 
\end{eqnarray}
Let us write the traceless field $h_{ab}^{(l)}$ as 
\begin{equation}
h^{(l)}_{ab} = h^{(r)}_{ab} + h^{(d)}_{ab}\,,  \label{trless}
\end{equation}
where
\begin{equation}
h^{(d)}_{ab}=\left(\nabla_a\nabla_b-\frac{3H^2}{\alpha-3}
g_{ab} \right)B\,.
\label{Bdef}
\end{equation}
The tensor $h^{(d)}_{ab}$ is traceless because of 
equation (\ref{eqnB}).  By taking the divergence of this equation twice,
we find 
\begin{equation}
\nabla^{a}\nabla^{b}h^{(d)}_{ab} 
= \frac{36\alpha H^4}{(\alpha-3)^2}B
= \nabla^{a}\nabla^{b}h^{(l)}_{ab}\,.  \label{deldelB}
\end{equation}
As a result we have $\nabla^a \nabla^b h_{ab}^{(r)} = 0$.  
Thus, the field
$h_{ab}$ can be decomposed as
\begin{equation}
h_{ab} = h_{ab}^{(r)} + h_{ab}^{(d)} + h_{ab}^{(t)}\,, \label{sectors}
\end{equation} 
where $h^{(d)}_{ab}$ and $h^{(t)}_{ab}$ are given by
(\ref{Bdef}) and (\ref{trdef}), respectively.  
We call the field $h^{(d)}_{ab} + h^{(t)}_{ab}$ the scalar sector
of $h_{ab}$ and the field $h^{(r)}_{ab}$ the non-scalar sector.  The latter
satisfies $\nabla^a \nabla^b h^{(r)}_{ab} = {h^{(r)a}}_a = 0$.

\section{Commutator functions of the scalar sector}

The contribution of the scalar sector to 
the two-point function can immediately be inferred once we find its 
contribution to the commutator (or Schwinger) function. For this reason
we calculate the latter in this section.

For any compactly-supported scalar function $f(x)$ we define 
\cite{WaldLectures}
\begin{eqnarray}
({\bf A}f)(x) & \equiv & \int dV_y G^{(S,A)}(x,y)f(y)\,, \\ 
({\bf R}f)(x) & \equiv  & \int dV_y G^{(S,R)}(x,y)f(y)\,,  \\ 
({\bf E}f)(x) & \equiv  & \int dV_y E^{(S)}(x,y)f(y) \nonumber  \\
                 & = & ({\bf A}f)(x) - ({\bf R}f)(x)\,. 
\end{eqnarray}
We define the advanced and retarded Green functions, 
$G^{(T,A)}_{aba'b'}(x,x')$ and $G^{(T,R)}_{aba'b'}(x,x')$, for the tensor
equation (\ref{Eulag1}) by requiring that
\begin{equation}
L^{(T)abcd}_xG^{(T,A/R)}_{cda'b'}(x,x')
=\frac{1}{2}\left(
\delta^a_{a'}\delta^b_{b'} + \delta^a_{b'}\delta^b_{a'}\right)\delta^4(x,x')\,,
\end{equation}
and that $G^{(T,A)}_{aba'b'}(x,x')$ ($G^{(T,R)}_{aba'b'}(x,x')$) vanish if
$x$ is in the future (past) of $x'$.  As in the scalar case we define the
advanced-minus-retarded Green function by
\begin{equation}
E^{(T)}_{aba'b'}(x,x') \equiv 
G^{(T,A)}_{aba'b'}(x,x') - G^{(T,R)}_{aba'b'}(x,x')\,.
\end{equation}
For any compactly-supported smooth tensor $h_{ab}(x)$ we define
$({\bf A}h)_{ab}(x)$, $({\bf R}h)_{ab}(x)$ and $({\bf E}h)_{ab}(x)$ in 
the same way as in the scalar case.
The equality shown by Peierls in \cite{Peierls} is quite general and holds
in this case as well.  Thus,
\begin{equation}
\left[ h_{ab}(x), h_{a'b'}(x')\right] = i E^{(T)}_{aba'b'}(x,x')\,.
\label{tensorcom}
\end{equation}
We will derive the commutator function for the scalar sector starting from this
formula.

Let us first examine the pure-trace part.
We obtain by a straightforward calculation 
\begin{equation}
L^{(T)abcd}\left(g_{cd}F\right)=
\frac{\alpha-3}{4}g^{ab}L^{(S)}F
\label{op1}
\end{equation}
for any smooth function $F$. 
Using this equation with $F={\bf R}f$, we have
\begin{eqnarray}
&& \int dV_x\,G^{(T,R)}_{efab}(y,x)L^{(T)abcd}_x\left[g_{cd}(x)
({\bf R}f)(x)\right] \nonumber \\
&& = \frac{\alpha-3}{4}\int dV_x\,G^{(T,R)}_{efab}(y,x)g^{ab}(x)f(x)\,.
\end{eqnarray}
We integrate by parts so that the
operator $L^{(T)cdab}_x$ is applied
on $G^{(T,R)}_{efab}(y,x)$ on the left-hand 
side.\footnote{There will be no boundary terms because
the common support of $G_{abcd}^{(T,R)}(y,x)$ 
and $({\bf R}f)(x)$ with fixed
$y$ is compact.}
Then we use
\begin{equation}
L^{(T)cdab}_x G^{(T,R)}_{efab}(y,x)
= \frac{1}{2}\left( \delta^c_e\delta^d_f + \delta^c_f \delta^d_e\right)
\delta^4(y,x)\,,
\end{equation}
which follows from the equality 
$G^{(T,A)}_{abcd}(x,y) = G^{(T,R)}_{cdab}(y,x)$.
Thus, we find
\begin{equation}
g_{ab}(y)\int dV_x G^{(S,R)}(y,x)f(x)
= \frac{\alpha-3}{4}\int dV_x\,G^{(T,R)}_{abcd}(y,x)
g^{cd}(x) f(x)\,.
\end{equation}
Since this holds for any compactly-supported smooth function $f(x)$, we
have
\begin{equation}
G^{(T,R)}_{abcd}(y,x)g^{cd}(x) = \frac{4}{\alpha-3}g_{ab}(y)G^{(S,R)}(y,x)\,.
\end{equation}
The same equality holds for the advanced Green functions
$G^{(T,A)}_{abcd}(y,x)$ and $G^{(S,A)}(y,x)$.  Therefore
\begin{equation}
E^{(T)}_{abcd}(x,y)g^{cd}(y) = \frac{4}{\alpha-3}g_{ab}(x)E^{(S)}(x,y)\,.
\label{Trpart}
\end{equation}
By taking the trace of equation (\ref{tensorcom}) and using (\ref{Trpart})
we find
\begin{equation}
\left[ h(x), h(x')\right] = i\,\frac{16}{\alpha-3} E^{(S)}(x,x')\,,
\label{trcom}
\end{equation}
where $h = g^{ab}h_{ab}$, and
\begin{eqnarray}
\left[ h_{ab}^{(r)}(x), h(x')\right] & = & 0\,, \label{com1} \\
\left[ B(x), h(x')\right] & = & 0\,, \label{trdcom}
\end{eqnarray}
where $h_{ab}^{(r)}$ and $B$ are defined in the previous section.

Next, we derive a relation similar to (\ref{Trpart}) for the traceless
part $h^{(d)}_{ab}$ in the scalar sector.  The key equation is 
\begin{equation}
L^{(T)abcd}\left(\nabla_c\nabla_d-\frac{1}{4}g_{cd}\Box\right)F=
\frac{3-\alpha}{4\alpha}\left(\nabla^a\nabla^b
-\frac{1}{4}g^{ab}\Box\right)L^{(S)}F\,, \label{KEY}
\end{equation}
which is valid for any smooth function $F$.
Useful identities in deriving this equation are
\begin{equation}
\Box\nabla_a\nabla_b F= 8H^2(\nabla_a\nabla_b-\frac{1}{4}g_{ab}\Box)F
+\nabla_a\nabla_b\Box F 
\end{equation}
and
\begin{equation}
\nabla_a\Box\nabla_b F = 3H^2\nabla_a\nabla_b F+ \nabla_a\nabla_b\Box F\,.
\label{id2}
\end{equation}
By a procedure similar to that which led to (\ref{Trpart}) we  
find from (\ref{KEY})
\begin{equation}
E^{(T)}_{aba'b'}(x,x')\left( \stackrel{\longleftarrow}{\nabla^{a'}}
\stackrel{\longleftarrow}{\nabla^{b'}}
-\frac{1}{4}\stackrel{\leftarrow}{\Box}_{x'}
g^{a'b'}(x')
\right)
= \frac{4\alpha}{3-\alpha}
\left(\nabla_a\nabla_b - \frac{1}{4}g_{ab}(x)\Box_x\right)
E^{(S)}(x,x') \,.
\end{equation}
The derivative operators with primed indices act at point $x'$ here
and in the rest of this paper.
By using this in (\ref{tensorcom}) we obtain
\begin{equation}
\left[ h_{ab}(x), B(x')\right] = 
i\, \frac{3-\alpha}{9H^4}\left[\nabla_a \nabla_b - 
\frac{1}{4}g_{ab}(x)\right]E^{(S)}(x,x')\,,
\end{equation}
where $B(x)$ is defined by (\ref{eqnB}). 
{}From this equation we find
\begin{eqnarray}
\left[ B(x),B(x')\right] = i\,\frac{3-\alpha}{9H^4}
E^{(S)}(x,x')\,,  \label{dcom} \\
\left[ h_{ab}^{(r)}(x),B(x')\right]
= 0\,, \label{com2} \\
\left[ h(x), B(x')\right] = 0\,. \label{trdcom2}
\end{eqnarray}

\section{Scalar sector of the graviton two-point function} 

The commutator functions found in the previous section
can be used to determine the commutators of
annihilation and creation operators in the scalar sector.
Since the fields $h(x)$ and $B(x)$ commute with
the non-scalar sector $h^{(r)}_{ab}(x)$,
the operators
$b_{l\sigma}$, $b_{l\sigma}^\dagger$, $c_{l\sigma}$ and $c_{l\sigma}^\dagger$
defined in (\ref{Bexp}) and (\ref{trexp})
commute with $h^{(r)}_{ab}(x)$. 
Also, equation (\ref{trdcom2}) shows that
the operators $b_{l\sigma}$ and $b_{l\sigma}^\dagger$ commute with
$c_{l\sigma}$ and $c_{l\sigma}^\dagger$.
Thus, the graviton two-point function can be written as
\begin{equation}
\langle 0|h_{ab}(x)h_{a'b'}(x')|0\rangle
= \Delta^{(r)}_{aba'b'}(x,x') + \Delta^{(s)}_{aba'b'}(x,x')\,,
\end{equation}
where 
\begin{eqnarray}
\Delta^{(r)}_{aba'b'}(x,x') & = & 
\langle 0|h^{(r)}_{ab}(x)h^{(r)}_{a'b'}(x')|0\rangle \,, \\
\Delta^{(s)}_{aba'b'}(x,x') & = & 
\langle 0|h^{(d)}_{ab}(x)h^{(d)}_{a'b'}(x')|0\rangle 
+ \langle 0|h^{(t)}_{ab}(x)h^{(t)}_{a'b'}(x')|0\rangle\,.
\end{eqnarray} 

The commutators of
annihilation and creation operators in the scalar sector 
can be found from equations (\ref{trcom}) and (\ref{dcom}) as
\begin{eqnarray}
\left[ b_{l'\sigma'}, b_{l\sigma}^\dagger\right] & = &  \frac{3-\alpha}{9H^4}
\delta_{ll'}\delta_{\sigma\sigma'}\,, \label{commu1}\\
\left[ c_{l'\sigma'}, c_{l\sigma}^\dagger\right] & = & \frac{16}{\alpha-3}
\delta_{ll'}\delta_{\sigma\sigma'} \label{commu2}
\end{eqnarray}
with all other commutators vanishing. 
Now, we require that
\begin{equation}
b_{l\sigma}|0\rangle = c_{l\sigma}|0\rangle = 0
\label{annihilate}
\end{equation}
for all $l$ and $\sigma$. 
Using the commutators (\ref{commu1}) and (\ref{commu2}), the expansion of
$B(x)$ and $h(x)$ in (\ref{Bexp}) and (\ref{trexp}) and the condition 
(\ref{annihilate}), we find
\begin{eqnarray}
\langle 0|B(x)B(x')|0\rangle  & = & \frac{3-\alpha}{9H^4}\Delta_+(x,x')\,,
 \\
\langle 0|h(x)h(x')|0\rangle  & = & \frac{16}{\alpha-3}\Delta_+(x,x')\,,
\end{eqnarray}
where $\Delta_+(x,x')$ is expressed in terms of the scalar mode functions
$\phi^{(l\sigma)}$ in (\ref{sctwo-pt}). 
{}From these equations and the expressions of $h^{(d)}_{ab}$ and
$h^{(t)}_{ab}$ in terms of $B$ and $h$, we immediately find
the scalar sector of the graviton two-point function as
\begin{eqnarray}
\Delta^{(s)}_{aba'b'}(x,x')  =  
\frac{3-\alpha}{9H^4}\left(\nabla_a\nabla_b-\frac{3H^2}{\alpha-3}g_{ab}\right)
\left(\nabla_{a'}\nabla_{b'}-\frac{3H^2}{\alpha-3}g_{a'b'}\right) 
\Delta_+(x,x')\nonumber \\
\ \ \  +\frac{16}{\alpha-3}\times
\frac{1}{16}g_{ab}g_{a'b'}\Delta_+(x,x') \nonumber \\
 = \left(\frac{3-\alpha}{9H^4}
\nabla_a\nabla_b\nabla_{a'}\nabla_{b'}+\frac{1}{3H^2}\nabla_a\nabla_b g_{a'b'}
+\frac{1}{3H^2}g_{ab}\nabla_{a'}\nabla_{b'}\right) \nonumber \\
\ \ \ \  \times \Delta_{+}(x,x')
\label{scalarcontribution}
\end{eqnarray}
with $g_{ab}=g_{ab}(x)$ and $g_{a'b'}=g_{a'b'}(x')$.
As we have seen, the function $\Delta_{+}(x,x')$ decreases at large distances
if $\alpha > 3$. Hence the scalar-sector two-point function 
$\Delta^{(s)}(x,x')$ decreases at large distances for these values
of $\alpha$.
Note also that each term in the scalar sector is pure gauge 
either at point $x$ or $x'$.\footnote{Related remarks were
made in \cite{Higuchi87:1} and
\cite{AllenTuryn}.} 
Hence, there will be no contribution 
from this sector to
the two-point function of a gauge-invariant quantity.
This is in agreement with the
expectation that the extra modes introduced in the
theory by gauge fixing should not contribute to physical quantities
at the tree level. 
In Appendix A 
we present an explicit form of the scalar-sector two-point function
for $\alpha=9$, where some simplification occurs.

Let us comment on the limit $\alpha \to 0$, i.e. the Landau gauge 
considered in \cite{AM}.  In this limit the field 
satisfies
\begin{equation}
\nabla^b h_{ab} = \frac{1}{4}\nabla_a h\,. \label{badcond}
\end{equation}
The two-point function $\Delta^{(s)}(x,x')$ remains in a
pure-gauge form in this limit.  However, the traceless part $h^{(l)}_{ab}$
defined by (\ref{trless}) satisfies 
$\nabla^a\nabla^b h^{(l)}_{ab} = 0$.
(As a result, some intermediate results such as equation (\ref{KEY})
are ill-defined.)  Therefore, if one imposed the condition
(\ref{badcond}) from the start 
without taking the $\alpha\to 0$ limit of the theory with nonzero $\alpha$,
the field would be decomposed as $h_{ab} = h^{(r)}_{ab} + h^{(t)}_{ab}$
with $\nabla^a\nabla^b h^{(r)}_{ab}=0$.  Therefore, the cancellation of 
physical contributions from the fields $h^{(d)}_{ab}$ and $h^{(t)}_{ab}$
could be overlooked.

We have not calculated the non-scalar sector two-point function,
$\Delta^{(r)}_{aba'b'}(x,x')$, 
which is of course necessary for finding the
full two-point function. 
The full two-point function can most easily be calculated by extending 
the work of Allen and Turyn~\cite{AllenTuryn} in the Euclidean 
approach. (They specialize to the choice $\alpha = 1$ and $\beta = -2$. 
Since the squared mass
of the modes in the scalar sector for this choice is $-6H^2$, 
their two-point function
grows badly at large distances in the scalar sector.) 
Our preliminary results with arbitrary values of $\alpha$ and $\beta$
show that it is impossible
to construct a covariant two-point
function which does not grow at large distances in this manner.  
However, the results in
\cite{HHT} and \cite{HigKou} imply that this growth in CGTF 
is also pure gauge.  We are currently investigating if this fact can be
verified directly. 

\begin{flushleft}
\large{\bf Acknowledgement}
\end{flushleft}
We thank Chris Fewster for useful discussions concerning the relation between
the symplectic product and the advanced-minus-retarded Green function and
Bernard Kay for helpful discussions on gauge invariant quantities.
We also thank Bruce Allen and Michael Turyn for useful communication.

\section*{Appendix A. The scalar sector of the two-point function
with $\alpha =9$}
\renewcommand{\theequation}{A\arabic{equation}}
\setcounter{equation}{0}

Note that for $\alpha =  9$ we have $12H^2/(\alpha-3) = 2H^2$.  Therefore
we have the conformally-coupled massless scalar field.
In this case the scalar two-point function (\ref{sctwo-pt}) is 
\begin{equation}
\Delta_+(x,x') 
=\frac{H^2}{16\pi^2}F(2,1;2;z) 
=\frac{H^2}{16\pi^2}\frac{1}{1-z}\,.
\end{equation}
The two-point function $\Delta^{(s)}_{aba'b'}(x,x')$
can be found from 
(\ref{scalarcontribution}) using~\cite{AllenJacobson}
\begin{eqnarray}
\nabla_a n_b & = & H\cot H\mu\,(g_{ab} - n_a n_b)\,,\\
\nabla_{a} n_{b'} & = & -\frac{H}{\sin H \mu}(g_{ab'} + n_a n_{b'})\,, 
\label{sin}\\
\nabla_{a} g_{bc'} & = & 
\frac{H(1-\cos H\mu)}{\sin H\mu}\,H(g_{ab}n_{c'}+g_{ac'}n_b)\,,
\end{eqnarray}
where
$n_a=\nabla_{a}\mu(x,x')$ is the tangent vector at point $x$
to the spacelike geodesic 
joining points $x$ and $x'$ (if there is such a spacelike geodesic).
The bi-vector $g_{ab'}$ is the
parallel propagator, i.e. for any vector $X^{a'}$ at point $x'$, 
$g_{ab'}X^{a'}$ is the vector at point $x$
obtained by parallelly transporting $X^{a'}$
along the geodesic.
The result is
\begin{eqnarray}
\Delta^{(s)}_{aba'b'}(x,x')  =  T_{1}(z)\,n_{a}n_{b}n_{a'}n_{b'}+
T_{2}(z)\,(g_{a'b'}n_{a}n_{b}+g_{ab}n_{a'}n_{b'}) \nonumber \\
\  +T_{3}(z)\,(g_{aa'}n_{b}n_{b'}+g_{ba'}n_{a}n_{b'}+
g_{ab'}n_{b}n_{a'}+g_{bb'}n_{a}n_{a'})
\nonumber \\
\  +T_{4}(z)\,(g_{ab'}g_{ba'}+g_{bb'}g_{aa'})+
T_{5}(z)\,g_{ab}g_{a'b'}\,, \label{explic}
\end{eqnarray}
where the coefficients $T_{i}(z)$ are given by
\begin{eqnarray}
T_{1}(z)& = & \frac{H^2}{24\pi^2}\left[
\frac{4}{z-1}
+\frac{24}{(z-1)^2}
+\frac{24}{(z-1)^3}
\right]\,, \\
T_{2}(z)& = & -\frac{H^2}{24\pi^2}\left[
\frac{1}{z-1}
+\frac{4}{(z-1)^2}
+\frac{3}{(z-1)^3}
 \right ]\,, \\
T_{3}(z) & = & \frac{H^2}{24\pi^2}\left [
\frac{2}{(z-1)^2}  
+\frac{3}{(z-1)^3}
\right]\,, \\
T_{4}(z) & = &  \frac{H^2}{24\pi^2}\left[
\frac{1}{2(z-1)^3}
\right]\,, \\
T_{5}(z) & = & \frac{H^2}{24\pi^2}\left[
\frac{1}{(z-1)^2}
+\frac{1}{2(z-1)^3}
\right]\,,
\end{eqnarray}
where $z=\cos^2\left( \frac{\mu(x,x')H}{2}\right)$.
As we have seen in section 2 the variable $z$ tends to $-\infty$
as the coordinate distance $r=\|{\bf x}-{\bf x}'\|$ in the coordinates
for metric (\ref{metric0}) tends to infinity.
In the present case 
the coefficients
$T_{i}$ tend to zero like $r^{-2}$. 
In the rest of this appendix we will calculate the components of
the tangent vector $n_a$ and the parallel propagator $g_{ab'}$ and see 
explicitly that
they are bounded as $r\to \infty$.  Thus, we will see that 
the two-point function
in (\ref{explic}) indeed tends to zero as $r\to \infty$.

Let us define
\begin{equation}
\chi(x,x') \equiv \cos H\mu  = 
\frac{\lam^2+{\lam'}^2-r^2}{2\lam\lam'}\,.
\end{equation}
We extend the function $\mu(x,x')$ with $x=(\lambda,{\bf x})$ and
$x' = (\lambda',{\bf x}')$ in the coordinates for metric
(\ref{metric0}) to the region with $\chi > 1$ by 
\begin{equation}
e^{iH\mu} = \chi + \sqrt{\chi^2 -1}\,.
\end{equation}
We then have
\begin{equation}
n_a = -\frac{i}{H\sqrt{\chi^2-1}}\nabla_a \chi
\end{equation}
and
\begin{equation}
\nabla_a \chi = - \frac{Hr}{\lambda'}V_a + H\left( \frac{\lambda}{\lambda'}
- \chi\right)t_a\,,
\end{equation}
where the vectors $V^a$ and $t^a$ at point $x=(\lambda,{\bf x})$
are defined by~\cite{Allen87:1} 
\begin{equation}
V^0  = 0\, \,,\ \ V^i = \frac{H\lambda (x^i - x^{\prime i})}{r}\,,
\end{equation}
with ``0" referring to the $\lambda$-component, and 
\begin{equation}
t^a = -H\lambda \left(\frac{\partial\ }{\partial \lambda}\right)^a\,.
\end{equation}
The unit vector $V^a$ is spacelike and the vector $t^a$ is a future-pointing
unit normal to the Cauchy surface with $\lambda = {\rm const}$, if we let
the variable $\lambda$ decrease towards the future.
(We define vectors $V^{a'}$ and $t^{a'}$ at $x'$ in a similar manner.
We note that $\lambda^{-1}V_{i} = - (\lambda')^{-1}V_{i'}$ in components.)
Thus, we obtain
\begin{equation}
n_{a}= \frac{ir}{\lambda'\sqrt{\chi^2 - 1}}V_a
- \frac{i}{\sqrt{\chi^2-1}}\left( \frac{\lambda}{\lambda'} - \chi\right)t_a\,.
\end{equation}
We note that, since $\chi$ grows like $r^2$ as $r$ increases,  
we have $n_a \to -it_a$ as $r\to \infty$.

The expression for the parallel propagator $g_{ab'}$ can be found from 
equation (\ref{sin}).
Thus, we have
\begin{equation}
g_{ab'} = \frac{1}{H^2}\nabla_{a}\nabla_{b'}\chi
- \frac{1}{H^2(\chi+1)}\nabla_a\chi\cdot\nabla_{b'}\chi\,.
\end{equation}
Let us define a bi-vector $P_{ab'}$ by $t^aP_{ab'} = t^{b'}P_{ab'} = 0$
and
\begin{equation}
P_{ij'} = \frac{1}{\lambda\lambda'H^2} \delta_{ij'}\,.
\end{equation}
Then 
\begin{equation}
H^{-2}\nabla_a \nabla_{b'}\chi
= P_{ab'} + \frac{r}{\lambda}t_a V_{b'} + \frac{r}{\lambda'}t_{b'}V_a 
+ \left(\chi - \frac{\lambda}{\lambda'} - \frac{\lambda'}{\lambda}\right)
t_a t_{b'}\,.
\end{equation}
Hence, the bi-vector $g_{ab'}$ can be written as
\begin{eqnarray}
g_{ab'} & = & P_{ab'} + \frac{1}{\chi+1}
\left( \chi - \frac{\lambda'}{\lambda} - \frac{\lambda}{\lambda'} -1\right)
t_a t_{b'} - \frac{r^2}{(\chi +1)\lambda\lambda'}V_aV_{b'}  \nonumber \\
&&  + \left(\frac{1}{\lambda} + \frac{1}{\lambda'}\right)\frac{r}{\chi + 1}
\left( t_a V_{b'} + t_{b'}V_a\right)\,.
\end{eqnarray}
We find
$g_{ab'} \to P_{ab'} + t_a t_{b'} + 2V_a V_{b'}$
as $r\to \infty$.

\end{document}